\documentclass[letterpaper, 10 pt, conference]{ieeeconf}  

\IEEEoverridecommandlockouts                              
\usepackage[dvips]{graphicx}
\usepackage{amsmath}
\newtheorem{theorem}{Theorem}
\newtheorem{remark}{Remark}
\newtheorem{lemma}{Lemma}
\newtheorem{definition}{Definition}
\newtheorem{assumption}{Assumption}

\begin{document}

\title{\LARGE \bf
Stochastic Aperiodic Control of Networked Systems\\ with i.i.d.\ Time-Varying Communication
Delays}


\author{Yohei Hosoe
\thanks{This work was partially supported by JST PRESTO Grant Number JPMJPR2127 and JSPS KAKENHI Grant Number 20K04546.}
\thanks{Y.~Hosoe is with the Department of Electrical Engineering, 
        Kyoto University, Nishikyo-ku, Kyoto 615-8510, Japan
        {\tt\small hosoe@kuee.kyoto-u.ac.jp}}%
}

\maketitle
\thispagestyle{empty}
\pagestyle{empty}

\begin{abstract}

This paper studies stochastic aperiodic stabilization of a networked control system (NCS) consisting of a
continuous-time plant and a discrete-time controller.
The plant and the controller are assumed to be connected by
communication channels with i.i.d.\ time-varying delays.
The delays are theoretically not required to be bounded even when the
plant is unstable in the deterministic sense.
In our NCS, the sampling interval is supposed to be determined directly by such
communication delays.
A necessary and sufficient inequality condition is presented for
designing a state-feedback controller stabilizing the NCS at
sampling points in a stochastic sense.
The results are also illustrated numerically.

\end{abstract}

\section{Introduction}

Communication delays are inevitable when transmitting signals via
communication channels.
Since the delays may affect performance of the networked control systems
(NCSs), 
taking them into account in analysis and synthesis is one of the
important issues in the field of NCSs \cite{zhang2017analysis,mahmoud2018fundamental,zhang2019networked}.
The delays naturally become randomly time-varying when
the Internet is used as the communication channels \cite{Paxson-IEEETN95}.
Hence, discrete-time stochastic processes (i.e., random sequences) are more desirable 
as the model of communication delays than constants in practice.

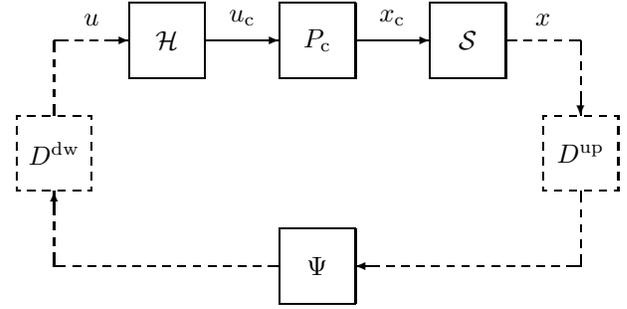
\begin{figure}[t]
	\centering
	\scalebox{1}{\begin{picture}(8,4)(0,0)
\put(1.5,3){\framebox(1,1){${\cal H}$}}
\put(2.5,3.5){\vector(1,0){1}}
\put(3,3.7){\makebox(0,0)[b]{$u_{\rm c}$}}
\put(3.5,3){\framebox(1,1){$P_{\rm c}$}}
\put(4.5,3.5){\vector(1,0){1}}
\put(5,3.7){\makebox(0,0)[b]{$x_{\rm c}$}}
\put(5.5,3){\framebox(1,1){${\cal S}$}}
\put(6.5,3.5){\line(1,0){0.05}} \multiput(6.55,3.5)(0.2,0){4}{\line(1,0){0.1}} \put(7.35,3.5){\line(1,0){0.15}}
\put(7,3.7){\makebox(0,0)[b]{$x$}}
\put(7.5,3.5){\line(0,-1){0.08}} \multiput(7.5,3.42)(0,-0.2){3}{\line(0,-1){0.1}} \put(7.5,2.82){\vector(0,-1){0.32}}
\put(7,1.5){\dashbox{0.1}(1,1){$D^{\rm up}$}}
\put(7.5,1.5){\line(0,-1){0.05}} \multiput(7.5,1.45)(0,-0.2){4}{\line(0,-1){0.1}} \put(7.5,0.65){\line(0,-1){0.15}}
\put(7.5,0.5){\line(-1,0){0.08}} \multiput(7.42,0.5)(-0.2,0){13}{\line(-1,0){0.1}} \put(4.82,0.5){\vector(-1,0){0.32}}
\put(3.5,0){\framebox(1,1){$\Psi$}}
\put(3.5,0.5){\line(-1,0){0.05}} \multiput(3.45,0.5)(-0.2,0){14}{\line(-1,0){0.1}} \put(0.65,0.5){\line(-1,0){0.15}}
\put(0.5,0.5){\line(0,1){0.08}} \multiput(0.5,0.58)(0,0.2){3}{\line(0,1){0.1}} \put(0.5,1.18){\vector(0,1){0.32}}
\put(0,1.5){\dashbox{0.1}(1,1){$D^{\rm dw}$}}
\put(0.5,2.5){\line(0,1){0.05}} \multiput(0.5,2.55)(0,0.2){4}{\line(0,1){0.1}} \put(0.5,3.35){\line(0,1){0.15}}
\put(0.5,3.5){\line(1,0){0.08}} \multiput(0.58,3.5)(0.2,0){3}{\line(1,0){0.1}} \put(1.18,3.5){\vector(1,0){0.32}}
\put(1,3.7){\makebox(0,0)[b]{$u$}}

\end{picture}}  
	\caption{Networked control system with random communication delays.}
	\label{fig:network}
\end{figure}

The structure of the NCS to be dealt with in this paper is standard
as in Fig.~\ref{fig:network}, where $P_{\rm c}$, $\Psi$, 
${\cal S}$ and ${\cal H}$ denote a continuous-time deterministic linear plant, a
discrete-time state-feedback controller, the ideal sampler and the zero-order
hold, respectively.
We denote this NCS by $\Sigma$.
The solid (resp. dashed) arrows are used for continuous-time
(resp. discrete-time) signals in the figure.
The discrete-time signals are transmitted via communication channels,
and the delay elements $D^{\rm up}$ and $D^{\rm
dw}$ are assumed to exist in the channels;
those elements delay the arrival of signals by some random intervals (the
details will be described later).
The sampler and the hold are assumed to operate in synchronization, 
which implies that the sampling intervals are determined by the random
delays; the sum of the two delays becomes the sampling
interval at each sampling time instant.
Hence, the arguments for stabilization of the NCS $\Sigma$ can be interpreted as a stochastic version of 
the aperiodic control \cite{hetel2017recent}.
The purpose of this paper is to show a synthesis-oriented inequality
condition for state-feedback stabilization of this type of NCSs in
the case that the delays (i.e., sampling intervals) are given by
i.i.d.\ processes (i.e., discrete-time white processes).

As can be observed from the above purpose,
our arguments are related with variable sampling\slash transmission
intervals and variable transmission delays, which are two of the five
types of imperfections and constraints in NCSs mentioned in \cite{mahmoud2018fundamental}.
The influence of the communication delays in NCSs has been
studied from various aspects since the article
\cite{anderson1989bilateral} was presented; see, e.g., the survey papers
\cite{zhang2017analysis,mahmoud2018fundamental,zhang2019networked} and
other sophisticated articles for details. 
In the case of internally stable plant $P_{\rm c}$,
it may be possible to ensure deterministic stability of the corresponding NCS
regardless of the boundedness of the delays, since a stable plant
without input is naturally stable.
In the case of unstable plant $P_{\rm c}$, on the other hand,
it is theoretically impossible to ensure deterministic stability of the corresponding NCS
under unbounded time-varying communication delays. 
Hence, if one desires to deal with unbounded communication delays in the
unstable case, some kind of ideas such as using
stochastic models are required.

If the delays are modeled by a stochastic process,
there is a room to deal with an unbounded support for the process even in
the unstable case \cite{montestruque2004stability,antunes2012volterra}.
Indeed, given a controller (which was designed for a virtual ideal NCS without
communication delays), the earlier study \cite{antunes2012volterra} discussed stability analysis of an
NCS in the presence of stochastic communication delays with unbounded
support.
To the best of the author's knowledge, however, there exist no earlier
 studies that enable us to design controllers directly (i.e., without
 relaxation) for NCSs in the
presence of stochastic communication delays with unbounded support;
the Markovian approaches, e.g., in \cite{zhang2005new,shi2009output} are also no exception
since the support of a finite-mode Markov chain is obviously bounded.
By exploiting the results in the author's earlier study \cite{Hosoe-TAC19}
about state feedback synthesis for discrete-time stochastic systems,
the present paper discusses synthesis of state-feedback controllers
stabilizing NCSs, which are optimal in the sense of stochastic stability
at sampling points.

Compared to the above earlier studies, 
the advantages of the proposed approach can be summarized as follows:
(i) the time-varying communication delays are not required to be
deterministically bounded; 
(ii) the distributions of the delays are arbitrary as long as they are
given by i.i.d.\ processes;
(iii) the derived synthesis-oriented inequality condition is necessary and sufficient in terms of
discrete-time stochastic stability (specifically, second-moment
exponential stability, which is also called mean square exponential stability);
(iv) the synthesis using the above condition is direct (i.e., no relaxation
is introduced).

This paper is organized as follows.
Section~\ref{sc:ncs} introduces NCSs with random communication delays to be dealt with in this paper.
Section~\ref{sc:dis} describes the discretized model of the NCSs, and states the associated controller synthesis problem.
Then, Section~\ref{sc:syn} discusses our main results about stabilizing controller synthesis.
For derivation of the inequality conditions about the synthesis, 
the results in \cite{Hosoe-TAC19} about stochastic systems are exploited.
Section~\ref{sc:exam} numerically illustrates our synthesis, and Section~\ref{sc:concl} gives concluding remarks.

We use the following notation in this paper.
The set of real numbers,
that of positive real numbers
and that of non-negative integers
are denoted by ${\bf R}$, ${\bf R}_+$ and ${\bf N}_0$, respectively.
The set of $n$-dimensional real
column vectors and that of $m\times n$ real matrices are denoted by
${\bf R}^n$ and ${\bf R}^{m\times n}$, respectively.
The set of 
$n\times n$ positive definite matrices is denoted by
${\bf S}^{n\times n}_{+}$.
The identity matrix of size $n$ is denoted by $I_n$. 
The Euclidean norm of
the vector $(\cdot)$
is denoted by $||(\cdot)||$.
The vectorization of the matrix $(\cdot)$ in the row
direction is denoted by ${\rm row}(\cdot)$, i.e., ${\rm row}(\cdot):=[{\rm row}_1(\cdot),\ldots,{\rm
row}_m(\cdot)]$, where $m$ is the number of rows of the matrix and ${\rm row}_i(\cdot)$ denotes the $i$th row.
The Kronecker product is denoted by $\otimes$.
The expectation (i.e., the expected value) of the random variable $(\cdot)$ is denoted by $E[(\cdot)]$; this notation
is also used for the expectation of a random matrix.
If $s$ is a random variable obeying the distribution ${\cal D}$,
then we represent it as $s \sim {\cal D}$.

\section{Networked Control Systems}
\label{sc:ncs}

\subsection{Settings of Components}

This subsection briefly states the settings of each of the components
constituting the NCS $\Sigma$ in Fig.~\ref{fig:network}.

\medskip\noindent
Plant $P_{\rm c}$: We suppose $P_{\rm c}$ is a continuous-time deterministic
linear plant represented by the state equation
\begin{align}
&
\dot{x}_{\rm c}(t)=A_{\rm c} x_{\rm c}(t) + B_{\rm c} u_{\rm c}(t),
\label{eq:con-plant}
\end{align}
where $t\in {\bf R}$ denotes the continuous time with initial time
$t=0$, and
$x_{\rm c}(t)\in {\bf R}^n$ and $u_{\rm c}(t)\in {\bf R}^m$ are the
state and the input, respectively.
The initial state $x_{\rm c}(0)=x_0$ is assumed to be given.

\medskip\noindent
Sampler ${\cal S}$ and Hold ${\cal H}$:
We suppose ${\cal S}$ and ${\cal H}$
are the sampler and the zero-order hold that operate in synchronization at sampling time instants
$t_k\ (k\in {\bf N}_0)$ satisfying
\begin{align}
&
t_0=0,\ \ t_{k+1}-t_k>0,\ \ \lim_{k\rightarrow \infty}t_k=\infty.\label{eq:sampling-inst}
\end{align}
We use the symbol $t$ for representing continuous time, and $k$ for
discrete time.
The sampling interval $h_k$ is defined as
\begin{align}
& h_k:=t_{k+1}-t_k. \label{eq:sampling-intr}
\end{align}
As is obvious from this definition, the sampling interval of our NCS is
not constant but time-varying.
The relation of the sampling interval with communication delays will be
described later.

\medskip\noindent
Controller $\Psi$: In this paper, we deal with a discrete-time
state-feedback controller for the NCS $\Sigma$.
For simplicity, we suppose that the controller $\Psi$ determines the control input $u_k$ immediately
after receiving the sampled output $x_k=x_{\rm c}(t_k)$ of the plant at
each discrete time $k$.
The class of the controller and its design method will be discussed later.

\medskip\noindent
Delays $D^{\rm up}$ and $D^{\rm dw}$:
We suppose $D^{\rm up}$ (resp. $D^{\rm dw}$) is a component that delays
(in the sense of continuous time)
the arrival of signal by $\tau_k^{\rm up}\geq 0$
(resp. $\tau_k^{\rm dw}\geq 0$) at each $k$.
Since the first timing of the sampling is $t=t_0=0$,
the controller $\Psi$ receives the plant output $x_{\rm c}(t_0)=x_0$ at $t=\tau_0^{\rm up}$.
Although the computation delay is not introduced in our study, it may be
included in $\tau_k^{\rm up}$ (or $\tau_k^{\rm dw}$) without loss of generality.
Since the output $u_0$ of $\Psi$ about $k=0$ is decided at
$t=\tau_0^{\rm up}$ in continuous time,
it reaches the hold ${\cal H}$ at $t=\tau_0^{\rm up}+\tau_0^{\rm dw}$.
Then, since the output of $P_{\rm c}$ is sampled simultaneously, $k$
changes from 0 to 1 at this timing, and hence, $t_1=\tau_0^{\rm
up}+\tau_0^{\rm dw}$.
This might be somewhat confusing since the hold ${\cal H}$ receives $u_0$ at
$t=t_1$, which implies that
\begin{align}
& u_{\rm c}(t)=u_0\label{eq:delayed-input0}
\end{align}
is added to $P_{\rm c}$ for $t\in[t_1, t_2)$.
The sampled plant output $x_1=x_{\rm c}(t_1)$ is transmitted to $\Psi$
via the communication channel again,
and the same process continues as $k$ increases.
As already stated, the sampler and the hold in our NCS operate in synchronization.
Hence, the relationship
\begin{align}
&
h_k=\tau_k^{\rm up}+\tau_k^{\rm dw} \label{eq:sampling-intr-delay}
\end{align}
holds for the sampling interval $h_k$ and the delays $\tau_k^{\rm up}$
and $\tau_k^{\rm dw}$.
For
\begin{align}
&
T_k^{\rm up}:=\sum_{\kappa=0}^{k}\tau_\kappa^{\rm up},\ \ 
T_k^{\rm dw}:=\sum_{\kappa=0}^{k}\tau_\kappa^{\rm dw},
\label{eq:sum-delay}
\end{align}
the sampling time instants are
\begin{align}
& t_0=0, t_k=T_{k-1}^{\rm
up}+T_{k-1}^{\rm dw}\ (k=1,2,\ldots).
\label{eq:sampling-inst-delay}
\end{align}
The timing of the decision of control input by $\Psi$ is
$t=T_{0}^{\rm up}, T_{1}^{\rm up}+T_{0}^{\rm dw}, T_{2}^{\rm up}+T_{1}^{\rm dw}, \ldots$.

\subsection{Stochastic Process Representation of Communication Delays}

As the models of time-varying delays $\tau_k^{\rm up}$ and $\tau_k^{\rm dw}$, this
paper consider stochastic processes.
We formally introduce the 2-dimensional stochastic process
\begin{align}
& \xi=\left(\xi_k\right)_{k\in {\bf N}_0},\ \ \xi_k=\begin{bmatrix}
	  \tau_k^{\rm up}& \tau_k^{\rm dw}
	  \end{bmatrix}^T.\label{eq:xi}
\end{align}
Then, only the following is used as the essential restriction on $\xi$ in this
paper.

\begin{assumption}
\label{as:iid}
$\xi_k$ is independent and identically distributed (i.i.d.) with
 respect to $k\in {\bf N}_0$.
\end{assumption}

This assumption does not deny the dependence between $\tau_k^{\rm up}$
and $\tau_k^{\rm dw}$ at common $k$.
The distribution of $\xi_k$ is arbitrary, and the support is not
required to be unbounded;
although another assumption will be additionally introduced on the NCS $\Sigma$ in the
following section, it is a minimal requirement for defining second-moment
stability, and does not imply the degradation of applicability of the
proposed approach.

\begin{remark}
Since the delays generally have physical lower bounds, we may also assume that
$\tau_k^{\rm up}$ and $\tau_k^{\rm dw}$ are positive in practice.
\end{remark}

\section{Discretization of NCS and Stabilization Problem}
\label{sc:dis}

We deal with the NCS $\Sigma$ in a discrete time domain (i.e., at
sampling points).
Under the aforementioned settings, the relationship between the
continuous-time signals and the discrete-time signals is given by
\begin{align}
&
x_k=x_{\rm c}(t_k),\ \ 
u_{\rm c}(t)=u_{k-1}\ \ 
(t\in [t_k, t_{k+1}), k\in {\bf N}_0).\label{eq:con-dis-signals}
\end{align}
With this relationship, the discretized model of the continuous-time plant
$P_{\rm c}$ can be given as
\begin{align}
& x_{k+1} = A(\xi_k) x_k + B(\xi_k) u_{k-1}, \label{eq:disc-plant}\\
&
A(\xi_k)=
e^{A_{\rm c}h_k},\ \ 
B(\xi_k)=
\int_0^{h_k}e^{A_{\rm c}t}B_{\rm c}dt\label{eq:dis-coefficient}
\end{align}
(recall (\ref{eq:sampling-intr-delay}) and (\ref{eq:xi})).
For this discretized model, we introduce the following assumption.
\begin{assumption}
\label{as:exp-bound}
The squares of entries of 
$A(\xi_k)$ and $B(\xi_k)$ in (\ref{eq:dis-coefficient})
are all Lebesgue integrable 
for $k=0$, i.e.,
\begin{align}
&
E[A_{ij}(\xi_0)^2]<\infty\ \ (i,j=1,\ldots,n),\\
&
E[B_{ij}(\xi_0)^2]<\infty\ \ (i=1,\ldots,n; j=1,\ldots,m),
\end{align}
where $A_{ij}(\xi_0)$ and $B_{ij}(\xi_0)$ are the $(i,j)$-entries of $A(\xi_0)$ and $B(\xi_0)$, respectively.
\end{assumption}

This assumption is a minimal requirement\footnote{This can be proved with Lemma~1 in \cite{Hosoe-TAC22}, although
details are omitted.} for defining second-moment
stability (in a discrete-time domain) for the present structure of NCS.
Hence, this assumption has the role of specifying the class of NCSs that can be
dealt with in the framework of second-moment stability;
if this assumption is not satisfied,
it is impossible to stabilize the NCS in the second moment.

As in (\ref{eq:disc-plant}), the discretized model has one step input delay,
which naturally stems from the communication delays.
Taking into account this input delay, we consider the following class of
state-feedback controller as $\Psi$ (the delayed input is dealt with
as part of the system state in the discrete time domain).%
\begin{align}
& u_{k} = F_1 x_k + F_2 u_{k-1}\label{eq:network-sf}
\end{align}
Then, the second-moment exponential stability of the NCS $\Sigma$ with
the plant (\ref{eq:con-plant}) and the controller (\ref{eq:network-sf})
can be defined as follows.

\begin{definition}
\label{df:expo}
Suppose that $\xi$ satisfies Assumption~\ref{as:iid}, and the random
matrices $A(\xi_k)$ and $B(\xi_k)$ in (\ref{eq:dis-coefficient}) satisfy
Assumption~\ref{as:exp-bound}.
For given $F_1\in {\bf R}^{m\times n}$ and $F_2\in {\bf R}^{m\times m}$,
the NCS $\Sigma$ is said to be exponentially stable
in the second moment at sampling points if there exist $a\in {\bf R}_+$ and $\lambda \in
 (0,1)$ such that
\begin{align}
&
\sqrt{E[||x_{\rm c}(t_k)||^2+||u_{\rm c}(t_{k})||^2]} \leq a
\lambda^k
\sqrt{||x_0||^2+||u_{-1}||^2}\notag\\
&
(\forall k \in {\bf
N}_0, \forall x_0 \in {\bf R}^n, \forall u_{-1} \in {\bf R}^m).
\label{eq:exp-def}
\end{align}
\end{definition}

In (\ref{eq:exp-def}), the expectation is taken for the random sampling time
instant $t_k$ given in (\ref{eq:sampling-inst-delay}).
The following section tackles the problem of designing $F_1$ and $F_2$
of $\Psi$ that stabilizes the corresponding NCS in the meaning of this definition.

\section{Synthesis of Stabilizing State-Feedback Controller}
\label{sc:syn}

Definition~\ref{df:expo} was introduced for the NCS
consisting of the continuous-time plant and the discrete-time
controller.
The inequality condition (\ref{eq:exp-def}) in the definition, however,
can be equivalently rewritten as that for the corresponding
discrete-time system, i.e., 
\begin{align}
&
\sqrt{E[||x_k||^2+||u_{k-1}||^2]} \leq a
\lambda^k
\sqrt{||x_0||^2+||u_{-1}||^2}\notag\\
&
(\forall k \in {\bf
N}_0, \forall x_0 \in {\bf R}^n, \forall u_{-1} \in {\bf R}^m),
\label{eq:exp-def-dist}
\end{align}
by using (\ref{eq:con-dis-signals}).
This implies that stabilization in the sense of Definition~\ref{df:expo} may be
achieved by designing a state-feedback controller (\ref{eq:network-sf})
that stabilizes the discretized model (\ref{eq:disc-plant}) by viewing it as a
standard discrete-time stochastic system (with one step input delay).
This section discusses such synthesis.

Let us formally introduce the new state variable
\begin{align}
& x_{{\rm e},k}=u_{k-1},\label{eq:art-delay}
\end{align}
and construct the extended system
\begin{align}
&
\hat{x}_{k+1}=\hat{A}(\xi_k) \hat{x}_k + \hat{B}u_k, \ \ 
\hat{x}_k=
\begin{bmatrix}
x_{k} \\ x_{{\rm e},k}
\end{bmatrix},\label{eq:ex-state}\\
&
\hat{A}(\xi_k)=
\begin{bmatrix}
A(\xi_k) & B(\xi_k) \\ 0 & 0
\end{bmatrix},\ \ 
\hat{B}=
\begin{bmatrix}
0 \\ I
\end{bmatrix}\label{eq:ex-coefficient}
\end{align}
from (\ref{eq:disc-plant}).
Obviously, the behavior of this system is consistent with that of the
discretized model (\ref{eq:disc-plant}).
With the state vector $\hat{x}_k$ of this extended system,
(\ref{eq:network-sf}) can also be rewritten as
\begin{align}
&
u_k = \hat{F}\hat{x}_k, \ \ \hat{F}=[F_1, F_2].\label{eq:ex-sf}
\end{align}
Then, we see that the discrete-time closed-loop system
\begin{align}
& 
\hat{x}_{k+1}=(\hat{A}(\xi_k) + \hat{B} \hat{F}) \hat{x}_k\label{eq:ex-cl}
\end{align}
consisting of
the extended system (\ref{eq:ex-state}) and the controller
(\ref{eq:ex-sf}) corresponds to the discrete-time counterpart of the
present NCS without loss of generality.

For such a closed-loop system,
(\ref{eq:exp-def-dist}) can be further equivalently rewritten as
\begin{align}
&
\sqrt{E[||\hat{x}_k||^2]} \leq a
\lambda^k
\sqrt{||\hat{x}_0||^2}\notag\\
&
(\forall k \in {\bf
N}_0, \forall \hat{x}_0 \in {\bf R}^{n+m}).
\label{eq:exp-def-dist-ext}
\end{align}
This is nothing but the inequality condition for defining second-moment
exponential stability of discrete-time stochastic systems
(see, e.g., Definition~2 in \cite{Hosoe-TAC19}).
 The explicit definition is given
as follows.
\begin{definition}
\label{df:expo-dist}
Suppose that $\xi$ satisfies Assumption~\ref{as:iid}, and the random
matrices $A(\xi_k)$ and $B(\xi_k)$ in (\ref{eq:dis-coefficient}) satisfy
Assumption~\ref{as:exp-bound}.
For given $\hat{F}\in {\bf R}^{m\times (n+m)}$,
the closed-loop system (\ref{eq:ex-cl}) is said to be exponentially stable
in the second moment if there exist $a\in {\bf R}_+$ and $\lambda \in
 (0,1)$ satisfying (\ref{eq:exp-def-dist-ext}).
\end{definition}

These observations lead us to the following lemma.
\begin{lemma}
\label{lm:equivalent}
Suppose that $\xi$ satisfies Assumption~\ref{as:iid}, and the random
matrices $A(\xi_k)$ and $B(\xi_k)$ in (\ref{eq:dis-coefficient}) satisfy
Assumption~\ref{as:exp-bound}.
For given $F_1\in {\bf R}^{m\times n}$ and $F_2\in {\bf R}^{m\times m}$,
the following two conditions are equivalent.
\begin{enumerate}
\item The NCS $\Sigma$ is exponentially stable in the second moment at
	  sampling points.
\item The discrete-time closed-loop system (\ref{eq:ex-cl}) is
	  exponentially stable in the second moment, where $\hat{A}(\xi_k)$,
	  $\hat{B}$ and $\hat{F}$ are given by (\ref{eq:ex-coefficient}) and (\ref{eq:ex-sf}).
\end{enumerate}
\end{lemma}

This lemma implies that the problem of designing a state-feedback
controller (\ref{eq:network-sf})
stabilizing the NCS reduces to that of designing a state-feedback
controller (\ref{eq:ex-sf})
stabilizing the closed-loop system (\ref{eq:ex-cl}).
Since the coefficient matrices (\ref{eq:ex-coefficient}) of the extended
system are depending only on $\xi_k$ (i.e., not on $\xi_{k-1},
\xi_{k-2}, \ldots$), the results in \cite{Hosoe-TAC19} can be used for
tackling the latter problem.
Hence, Lemma~\ref{lm:equivalent}, together with the results in \cite{Hosoe-TAC19}, leads us to the
following theorems about stability analysis and synthesis of NCSs, which
constitute a part of the main results in this paper.

\begin{theorem}
\label{th:anal}
Suppose that $\xi$ satisfies Assumption~\ref{as:iid}, and the random
matrices $A(\xi_k)$ and $B(\xi_k)$ in (\ref{eq:dis-coefficient}) satisfy
Assumption~\ref{as:exp-bound}.
For given $F_1\in {\bf R}^{m\times n}$ and $F_2\in {\bf R}^{m\times m}$,
the NCS $\Sigma$ is exponentially stable in the second moment at sampling points
if and only if there exist $P\in {\bf S}^{(n+m)\times (n+m)}_+$ and $\lambda\in
(0,1)$ satisfying
\begin{align}
&
\lambda^2 P - \widetilde{G}^T (P\otimes I_{\bar{n}}) \widetilde{G} \geq 0,\label{eq:lmi-anal}
\end{align}
where $\widetilde{G}$ is the matrix given by
\begin{align}
&
\widetilde{G} :=[\bar{G}_{1}^T, \ldots,
 \bar{G}_{n+m}^T]^T \in {\bf R}^{(n+m)\bar{n} \times (n+m)},\label{eq:def-Gp}\\
&
\bar{G}=:
\left[\bar{G}_{1}, \ldots, \bar{G}_{n+m}\right] \notag\\ 
&
(\bar{G}_{i} \in {\bf R}^{\bar{n}\times (n+m)}\ (i=1,\ldots,n+m))
\end{align}
with a matrix $\bar{G} \in {\bf R}^{\bar{n}\times(n+m)^2}\ (\bar{n}\leq (n+m)^2)$ satisfying
\begin{align}
&
\bar{G}^T \bar{G}=E\big[{\rm row}(\hat{A}(\xi_0)+\hat{B}\hat{F})^T
{\rm row}(\hat{A}(\xi_0)+\hat{B}\hat{F})\big]
\label{eq:equiv-rep-decom-syn}
\end{align}
for $\hat{A}(\xi_k)$, $\hat{B}$ and $\hat{F}$ in
(\ref{eq:ex-coefficient}) and (\ref{eq:ex-sf}). 
\end{theorem}

\begin{theorem}
\label{th:syn}
Suppose that $\xi$ satisfies Assumption~\ref{as:iid}, and the random
matrices $A(\xi_k)$ and $B(\xi_k)$ in (\ref{eq:dis-coefficient}) satisfy
Assumption~\ref{as:exp-bound}.
There exists a gain $[F_1, F_2]=\hat{F}\in {\bf R}^{m\times (n+m)}$ such that the NCS $\Sigma$
is exponentially stable in the second moment at sampling points
if and only if there exist $X\in {\bf S}^{(n+m)\times (n+m)}_+$,
$Y\in {\bf R}^{m\times (n+m)}$ and $\lambda\in (0,1)$ satisfying
\begin{align}
&
\begin{bmatrix}
\lambda^2 X& \ast\\
\widetilde{G}_{A}X+\widetilde{G}_{B} Y &
X\otimes I_{\bar{m}}
\end{bmatrix}> 0, \label{eq:lmi-syn}
\end{align}
where $\widetilde{G}_{A}$ and $\widetilde{G}_{B}$ are the matrices given by
\begin{align}
&
\widetilde{G}_A :=[\bar{G}_{A1}^T, \ldots,
 \bar{G}_{A(n+m)}^T]^T \in {\bf R}^{(n+m)\bar{m} \times (n+m)},\label{eq:def-GpA}\\
&
\widetilde{G}_B :=[\bar{G}_{B1}^T, \ldots,
\bar{G}_{B(n+m)}^T]^T \in {\bf R}^{(n+m)\bar{m} \times m},
\label{eq:def-GpB}\\
&
\bar{G}_{AB}=:
\left[\bar{G}_{A1}, \ldots, \bar{G}_{A(n+m)}, \bar{G}_{B1}, \ldots,
\bar{G}_{B(n+m)}\right] \notag\\
& 
(\bar{G}_{Ai} \in {\bf R}^{\bar{m}\times (n+m)}, \bar{G}_{Bi} \in {\bf
R}^{\bar{m}\times m}\ (i=1,\ldots,n))
\end{align}
with a matrix $\bar{G}_{AB} \in {\bf R}^{\bar{m}\times(n+m)(n+2m)}\ (\bar{m}\leq (n+m)(n+2m))$ satisfying
\begin{align}
&
\bar{G}_{AB}^T \bar{G}_{AB}=E\big[[{\rm row}(\hat{A}(\xi_0)), {\rm row}(\hat{B})]^T
\notag \\
&\hspace{15mm}\cdot [{\rm row}(\hat{A}(\xi_0)), {\rm row}(\hat{B})]\big]
\label{eq:equiv-rep-decom-syn}
\end{align}
for $\hat{A}(\xi_k)$, $\hat{B}$ and $\hat{F}$ in
(\ref{eq:ex-coefficient}) and (\ref{eq:ex-sf}). 
In particular, $[F_1, F_2]=\hat{F}=YX^{-1}$ is one such stabilizing gain.
\end{theorem}

Theorem~\ref{th:anal} is about analysis, while Theorem~\ref{th:syn} is about synthesis.
By calculating $\widetilde{G}$, for fixed $\lambda$, (\ref{eq:lmi-anal}) can be solved as a standard linear matrix
inequality (LMI).
A similar comment also applies to (\ref{eq:lmi-syn}) about synthesis.

\begin{remark}
The matrices $\widetilde{G}_A$ and $\widetilde{G}_B$ in (\ref{eq:lmi-syn}) are constructed so that
\begin{align}
&
\widetilde{G}_A^T (X\otimes I_{\bar{m}}) \widetilde{G}_A = E\left[\hat{A}(\xi_0)^T X \hat{A}(\xi_0)\right],\\
&
\widetilde{G}_A^T (X\otimes I_{\bar{m}}) \widetilde{G}_B = E\left[\hat{A}(\xi_0)^T X \hat{B}\right],\\
&
\widetilde{G}_B^T (X\otimes I_{\bar{m}}) \widetilde{G}_B = \hat{B}^T X \hat{B}.
\end{align}
are simultaneously satisfied.
This is an essence of deriving LMI conditions for stochastic systems.
In the case of the present networked control system, however, taking $\bar{G}_{AB}$ 
as in (\ref{eq:equiv-rep-decom-syn}) may actually be redundant,
 because $\hat{B}$ is known to be a constant matrix.
Indeed, it can be shown with the results in \cite{Hosoe-Auto20} that taking $\widetilde{X}_{1A}\in {\bf R}^{\bar{m}\times(n+m)^2 + 1}$ satisfying
$\bar{X}_{1A}^T \bar{X}_{1A} = E\big[[1, {\rm row}(\hat{A}(\xi_0))]^T  [1,{\rm row}(\hat{A}(\xi_0))]\big]$
is sufficient for constructing $\widetilde{X}_A \in {\bf R}^{(n+m)\bar{m} \times (n+m)}$ and $\widetilde{X}_I \in {\bf R}^{(n+m)\bar{m} \times (n+m)}$ such that
\begin{align}
&
\widetilde{X}_A^T (X\otimes I_{\bar{m}}) \widetilde{X}_A = E\left[\hat{A}(\xi_0)^T X \hat{A}(\xi_0)\right],\\
&
\widetilde{X}_A^T (X\otimes I_{\bar{m}}) \widetilde{X}_I = E\left[\hat{A}(\xi_0)^T X \right],\\
&
\widetilde{X}_I^T (X\otimes I_{\bar{m}}) \widetilde{X}_I = X.
\end{align}
With such matrices,
 $\widetilde{G}_A$ and $\widetilde{G}_B$ in (\ref{eq:lmi-syn}) can be replaced by $\widetilde{X}_A$ and $\widetilde{X}_I \hat{B}$, respectively.
Since the size of $\bar{X}_{1A}$ is generally smaller than that of $\bar{G}_{AB}$, this idea would contribute to reducing the cost for computing the expectations.
\end{remark}

\section{Numerical Example}
\label{sc:exam}
Let us consider the continuous-time deterministic linear plant (\ref{eq:con-plant}) with 
coefficient matrices
\begin{align}
A_{\rm c}=
\begin{bmatrix} 
0 & 1\\ \frac{g}{r} & 0
\end{bmatrix},\ \ 
B_{\rm c}=
\begin{bmatrix} 
0 \\ \frac{1}{Mr^2}
\end{bmatrix}
\end{align}
as $P_{\rm c}$ in Fig.~\ref{fig:network},
where $g=9.8$, $r=0.2$ and $M=1$.
This plant corresponds to a linearized model of an inverted pendulum around the vertically inverted position ($x_{{\rm c}1}$, $x_{{\rm c}2}$ and $u_{\rm c}$ correspond to the angle $\theta$, the angular velocity $\dot{\theta}$ and the input torque $\tau$, respectively).
Hence, the plant itself is unstable.
This section deals with an example of designing a state-feedback controller $\Psi$ stabilizing this plant from a remote location.
The communication delays $\tau^{\rm up}_k$ and $\tau^{\rm dw}_k$ are assumed to be i.i.d.\ (i.e., Assumption~\ref{as:iid} is satisfied) and given by
\begin{align}
&
\tau^{\rm up}_k=\epsilon^{\rm up}+d^{\rm up}_k,\ \ \tau^{\rm dw}_k=\epsilon^{\rm dw}+d^{\rm dw}_k,
\label{eq:tau-ed}\\
&
d^{\rm up}_k \sim {\rm Exp}(\mu^{\rm up}),\ \ d^{\rm dw}_k \sim {\rm Exp}(\mu^{\rm dw}),
\label{eq:d-exp}\\
&
\epsilon^{\rm up} = \epsilon^{\rm dw} = 0.01, \ \ \mu^{\rm up}=0.01,\ \ \mu^{\rm dw}=0.02,
\label{eq:ddmm}
\end{align}
where ${\rm Exp}(\mu)\ (\mu>0)$ denotes the exponential distribution with mean $\mu$.
For simplicity, $\tau^{\rm up}_k$ and $\tau^{\rm dw}_k$ are also assumed to be independent of each other at common $k$.
With these problem settings, Assumption~\ref{as:exp-bound} can be confirmed to be satisfied (see Appendix).
Hence, our approach can be applied to the corresponding NCS $\Sigma$.

To design a controller based on Theorem~\ref{th:syn}, we first have to construct the matrices $\widetilde{G}_A$ and $\widetilde{G}_B$.
Since the transformation from $\bar{G}_{AB}$ to those matrices is elementary, the only potentially tricky part is the computation about (\ref{eq:equiv-rep-decom-syn}).
In this paper, we generated 1,000 samples of $\xi_0=[\tau^{\rm up}_0, \tau^{\rm dw}_0]^T$ with the aforementioned exponential distributions, and replaced the expectation in (\ref{eq:equiv-rep-decom-syn}) by the corresponding sample mean\footnote{This approach allows us to use measured values of communication delays directly, which are the samples generated from distributions god only knows.}.
We used MATLAB and Statistics and Machine Learning Toolbox for such sample generation.
Then, we decomposed the matrix-valued sample mean as in (\ref{eq:equiv-rep-decom-syn}) by using 
the singular value decomposition.
The minimal value of $\bar{m}$, which corresponds to the rank of the matrix-valued sample mean, was 3.
Hence, the sizes of $\widetilde{G}_A$ and $\widetilde{G}_B$ became $9\times3$ and $9\times 1$, respectively.

With those coefficient matrices, we solved the LMI (\ref{eq:lmi-syn}) so that a minimal $\lambda$ is obtained through a bisection method.
Then, we obtained $\lambda=0.7628$ and the corresponding solution
\begin{align}
&
X=10^4
\begin{bmatrix}
   0.1854 &  -1.2981   & 0.0000\\
   -1.2981 &   9.0869  & -0.0002\\
    0.0000  & -0.0002 &   0.0001
\end{bmatrix}, \notag \\
&
Y=
\begin{bmatrix}
0.1003  & -1.7149 &   0.2836
\end{bmatrix},
\end{align}
which leads to 
\begin{align}
&
F_1=
\begin{bmatrix}
-5.5264 &  -0.7895  
\end{bmatrix}, \ \ 
F_2=-0.8488.
\end{align}
We used MATLAB, YALMIP \cite{YALMIP} and SDPT3 \cite{SDPT3} for this computation.

\begin{figure}[t] 
   \centering
      \includegraphics[width=3.5in]{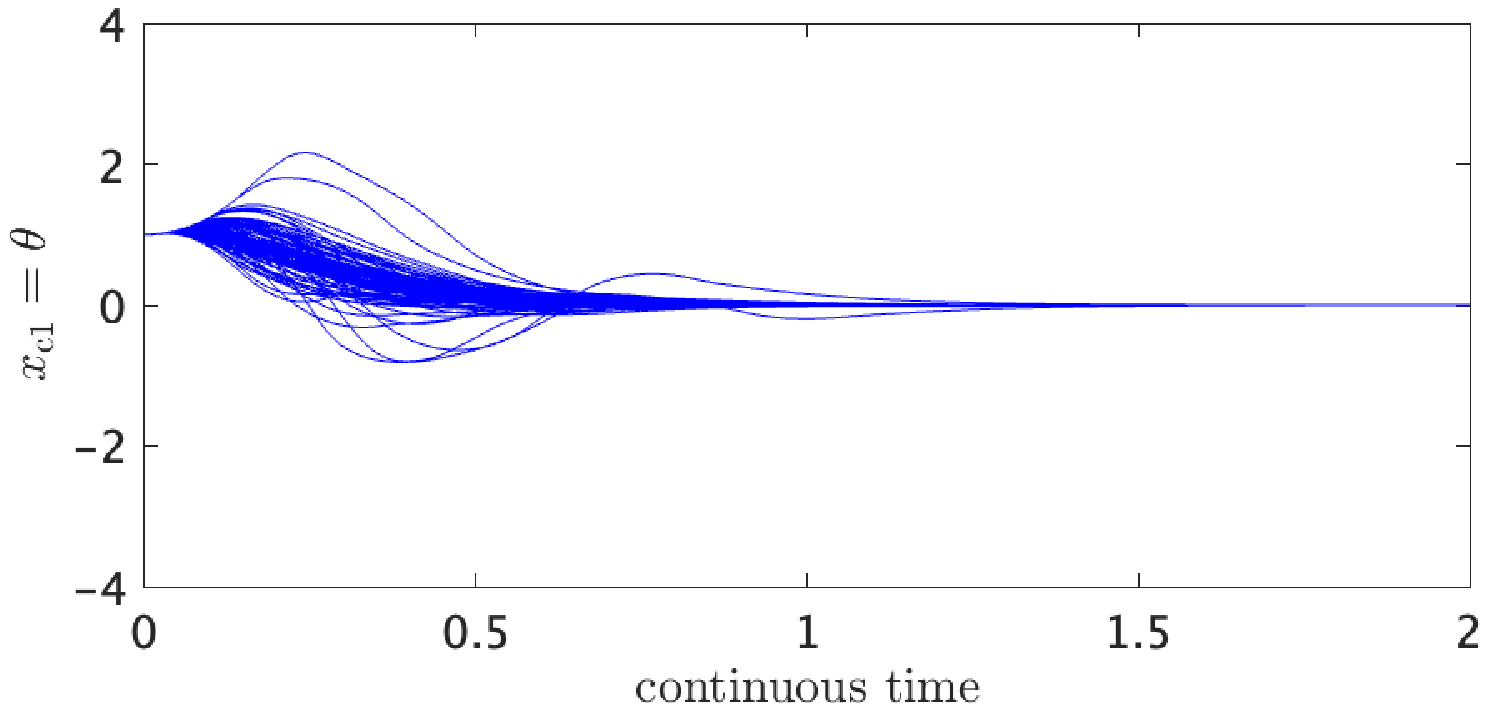} 
         \includegraphics[width=3.5in]{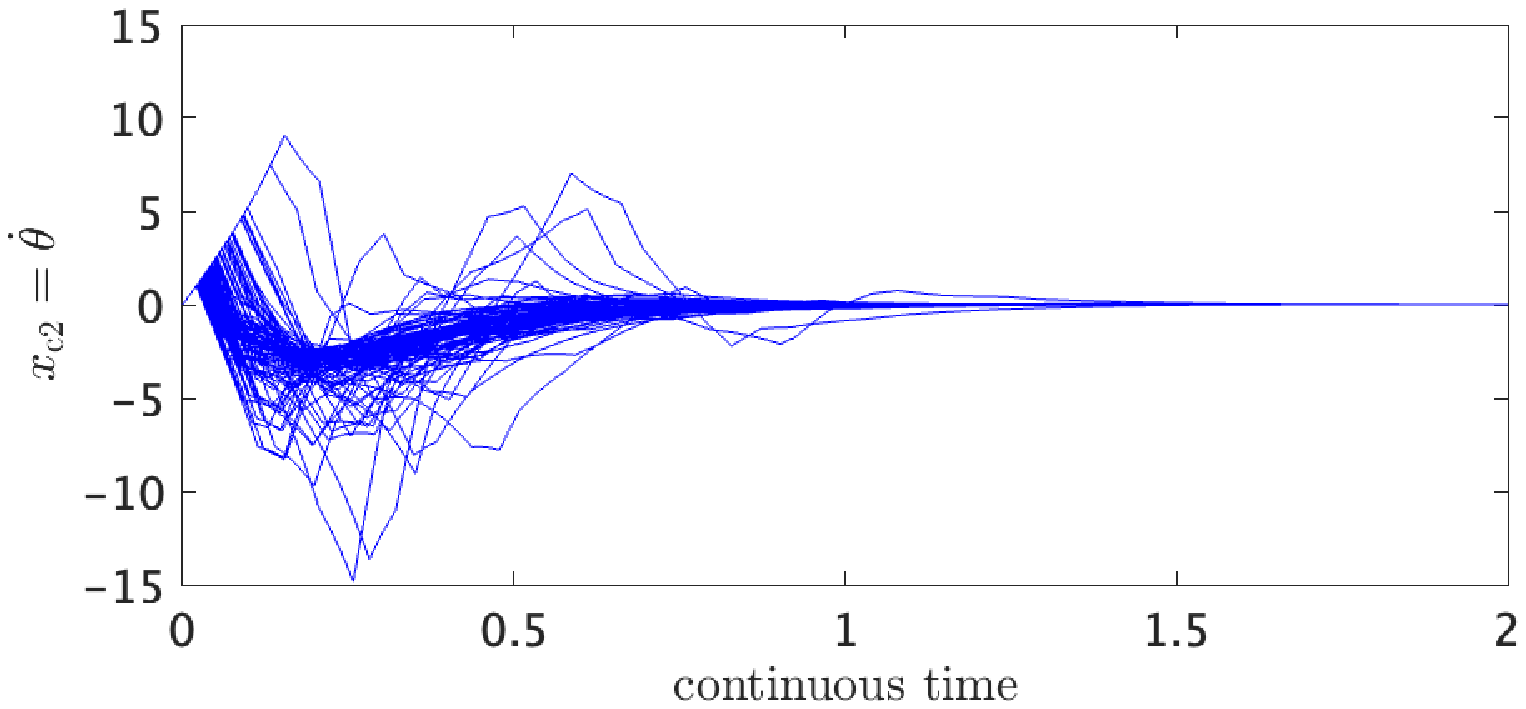} 
   \includegraphics[width=3.5in]{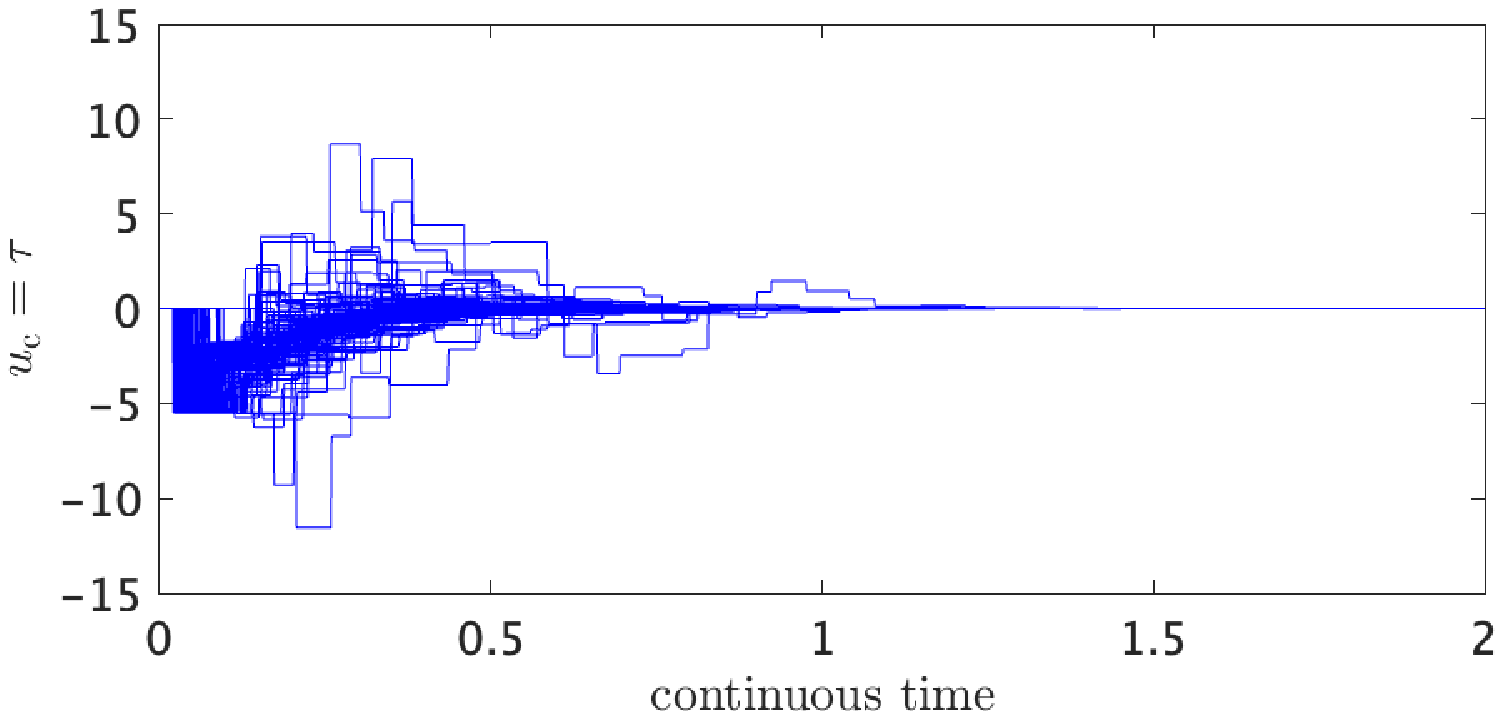} 
   \caption{Initial value response of NCS with designed gains.}
   \label{fig:initial}
\end{figure}

Since $\lambda < 1$, the NCS $\Sigma$ with obtained $F_1$ and $F_2$ is ensured to be stable by our results.
This can be also confirmed through the initial value response of the NCS shown in Fig.~\ref{fig:initial},
where 100 sample paths generated under the initial condition $x_c(0)=[1, 0]^T$ and $u_{-1}=0$ are overlapped.
All the sample paths in this figure converge to zero as time tends to infinity, even with the random communication delays whose behavior can be imagined from the sample paths about $u_{\rm c}$.
This kind of control is difficult to achieve only with deterministic approaches, since the plant is unstable and the range of the communication delays is not bounded.

\section{Conclusions}
\label{sc:concl}
In this paper, we tackled a stabilization problem for NCSs with time-varying random communication delays,  and proposed numerically tractable necessary and sufficient inequality conditions.
With our approach, the plant in the NCS itself is not required to be deterministically stable even when the range of the random communication delays is not (known to be) bounded.
The distribution of the delays may be constructed directly from data, and the approach is also considered to be compatible with some learning approaches in this sense.

\section*{Appendix}

We show that Assumption~\ref{as:exp-bound} is satisfied in the case of the present numerical example.
The assertion about the $A$ matrix in the assumption is equivalent to the boundedness of the maximum eigenvalue of 
the positive semidefinite matrix $E[A(\xi_0)^T A(\xi_0)]$ (recall the definition of maximum singular values).
Hence, we prove each entry of this matrix is bounded.

Since a direct calculation leads to
\begin{align}
&
E[A(\xi_0)^T A(\xi_0)] \notag\\
 =&
E\left[e^{A_{\rm c}^T(\epsilon^{\rm up}+\epsilon^{\rm dw}+d^{\rm up}_0+d^{\rm dw}_0)} e^{A_{\rm c}(\epsilon^{\rm up}+\epsilon^{\rm dw}+d^{\rm up}_0+d^{\rm dw}_0)}\right]\notag\\
=&
e^{A_{\rm c}^T(\epsilon^{\rm up}+\epsilon^{\rm dw})}
E\left[e^{A_{\rm c}^T(d^{\rm up}_0+d^{\rm dw}_0)} e^{A_{\rm c}(d^{\rm up}_0+d^{\rm dw}_0)}\right]\notag\\
&
\cdot
e^{A_{\rm c}(\epsilon^{\rm up}+\epsilon^{\rm dw})},
\end{align}
it is sufficient to show the boundedness of each entry of $E\left[e^{A_{\rm c}^T(d^{\rm up}_0+d^{\rm dw}_0)} e^{A_{\rm c}(d^{\rm up}_0+d^{\rm dw}_0)}\right]$.
For simplicity, we use the diagonalizability of $A_{\rm c}$.
Let $\alpha=\sqrt{g/r}\ (>0)$.
Then, the matrix $A_{\rm c}$ in the example has real eigenvalues $\lambda_1=\alpha$ and $\lambda_2=-\alpha$,
and can be diagonalized with the matrix 
$V:=\begin{bmatrix} 1 & -1\\ \alpha & \alpha \end{bmatrix}$
as
\begin{align}
&
V^{-1} A_{\rm c} V = \Lambda :=
\begin{bmatrix} \lambda_1  & 0\\  0 & \lambda_2 \end{bmatrix}.
\end{align}
This leads us to
\begin{align}
&
E\left[e^{A_{\rm c}^T(d^{\rm up}_0+d^{\rm dw}_0)} e^{A_{\rm c}(d^{\rm up}_0+d^{\rm dw}_0)}\right] \notag \\
=&
E\left[ V^{-T} e^{\Lambda(d^{\rm up}_0+d^{\rm dw}_0)}V^T V e^{\Lambda(d^{\rm up}_0+d^{\rm dw}_0)} V^{-1}\right] \notag \\
=&
V^{-T}E\left[e^{\Lambda d^{\rm dw}_0}e^{\Lambda d^{\rm up}_0}V^T V e^{\Lambda d^{\rm up}_0}e^{\Lambda d^{\rm dw}_0}\right] V^{-1} \notag \\
=&
V^{-T} E\left[e^{\Lambda d^{\rm dw}_0}E[e^{\Lambda d^{\rm up}_0}V^T V e^{\Lambda d^{\rm up}_0}]e^{\Lambda d^{\rm dw}_0}\right] V^{-1},
\end{align}
where the third equality follows from the independence of $d^{\rm up}_0$ and $d^{\rm dw}_0$.
Since $V^T V$ is a constant matrix,
each entry of $E[e^{\Lambda d^{\rm up}_0}V^T V e^{\Lambda d^{\rm up}_0}]$ is bounded if 
$E[e^{(\lambda_i+\lambda_j) d^{\rm up}_0}]$ is bounded for all $i,j=1,2$.
Similarly, in the case that $E[e^{\Lambda d^{\rm up}_0}V^T V e^{\Lambda d^{\rm up}_0}]$ is a bounded constant matrix, 
each entry of $E\left[e^{\Lambda d^{\rm dw}_0}E[e^{\Lambda d^{\rm up}_0}V^T V e^{\Lambda d^{\rm up}_0}]e^{\Lambda d^{\rm dw}_0}\right]$ 
is bounded if  $E[e^{(\lambda_i+\lambda_j) d^{\rm dw}_0}]$ is bounded for all $i,j=1,2$.
We can confirm through a direct calculation that 
$E[e^{(\lambda_i+\lambda_j) d^{\rm up}_0}]$ and $E[e^{(\lambda_i+\lambda_j) d^{\rm dw}_0}]$
are bounded for all $i,j=1,2$ in the present example; e.g.,
\begin{align}
E[e^{(\lambda_1+\lambda_1) d^{\rm dw}_0}]
&=
\frac{1}{\mu^{\rm dw}}
\int_0^\infty e^{(2\lambda_1-\frac{1}{\mu^{\rm dw}}) t}dt \notag \\
&= {\rm const.},
\end{align}
since $2\lambda_1-\frac{1}{\mu^{\rm up}}=-36<0$.
Therefore, 
each entry of $E\left[e^{A_{\rm c}^T(d^{\rm up}_0+d^{\rm dw}_0)} e^{A_{\rm c}(d^{\rm up}_0+d^{\rm dw}_0)}\right]$ is bounded,
and the assertion about the $A$ matrix in Assumption~\ref{as:exp-bound} is satisfied.

The above arguments about the $A$ matrix, together with the direct calculation of the integral in (\ref{eq:dis-coefficient}), also imply that the assertion about the $B$ matrix in Assumption~\ref{as:exp-bound} is satisfied.
Hence, Assumption~\ref{as:exp-bound} is satisfied.


\bibliographystyle{IEEEtran}
\bibliography{rds}

%
%
%
%

\end{document}